\documentclass[aps,prx,twocolumn,notitlepage,showpacs,amsmath,amstex,amssymb,citeautoscript,longbibliography]{revtex4-1}
\pdfoutput=1
\usepackage[english]{babel}
\usepackage{letltxmacro}
\usepackage{latexsym}
\usepackage{appendix}
\LetLtxMacro{\ORIGselectlanguage}{\selectlanguage}
\makeatletter
\DeclareRobustCommand{\selectlanguage}[1]{%
  \@ifundefined{alias@\string#1}
    {\ORIGselectlanguage{#1}}
    {\begingroup\edef\x{\endgroup
       \noexpand\ORIGselectlanguage{\@nameuse{alias@#1}}}\x}%
}
\newcommand{\definelanguagealias}[2]{%
  \@namedef{alias@#1}{#2}%
}
\makeatother
\definelanguagealias{en}{english}
\definelanguagealias{English}{english}
\usepackage{graphicx}
\usepackage{amsmath}
\usepackage{amsfonts}
\usepackage{amssymb,bbding}
\usepackage{bm}
\usepackage{color}
\usepackage[percent]{overpic}
\usepackage{esvect}
\usepackage{soul} 
\usepackage{amssymb}
\usepackage{wasysym}
\usepackage{dsfont}
\usepackage{float}
\usepackage{braket}
\usepackage{cancel}
\usepackage{comment}
\usepackage{mwe}
\usepackage{mathtools}

\usepackage{hyperref}
\hypersetup{
    bookmarks=false,         
    unicode=false,          
    pdftoolbar=false,        
    pdfmenubar=true,        
    pdffitwindow=false,     
    pdfstartview={FitH},    
    pdftitle={},    
    pdfauthor={},     
    pdfsubject={},   
    pdfcreator={},   
    pdfproducer={}, 
    pdfkeywords={quantum scars} {non-ergodic dynamics}, 
    pdfnewwindow=true,      
    colorlinks=true,       
    linkcolor=black,          
    citecolor=blue,        
    filecolor=magenta,      
    urlcolor=blue           
}

\newcommand{\be}{\begin{equation}}
\newcommand{\ee}{\end{equation}}
\newcommand{\bea}{\begin{eqnarray}}
\newcommand{\eea}{\end{eqnarray}}

\newcommand{\tr}{\mathop{\rm tr}}

\newcommand{\bigzero}{\mbox{\normalfont\Large\bfseries 0}}
\newcommand{\rvline}{\hspace*{-\arraycolsep}\vline\hspace*{-\arraycolsep}}

\newcommand{\papername}{Area-law entangled eigenstates from nullspaces of local Hamiltonians}
\begin{document}
\author{Volker Karle}
\affiliation{IST Austria, Am Campus 1, 3400 Klosterneuburg, Austria}
\author{Maksym Serbyn}
\affiliation{IST Austria, Am Campus 1, 3400 Klosterneuburg, Austria}
\author{Alexios A. Michailidis}
\affiliation{IST Austria, Am Campus 1, 3400 Klosterneuburg, Austria}
\title{\papername}
\begin{abstract}
Eigenstate thermalization in quantum many-body systems implies that eigenstates at high energy are similar to random vectors. Identifying systems where at least some eigenstates are non-thermal is an outstanding question. In this work we show that interacting quantum models that have a nullspace --- a degenerate subspace of eigenstates at zero energy (zero modes), which corresponds to infinite temperature, provide a route to non-thermal eigenstates.  We analytically show the existence of a zero mode which can be represented as a matrix product state for a certain class of local Hamiltonians. In the more general case we use a subspace disentangling algorithm to generate an orthogonal basis of zero modes characterized by increasing entanglement entropy. We show evidence for an area-law entanglement scaling of the least entangled zero mode in the broad parameter regime, leading to a conjecture that all local Hamiltonians with the nullspace  feature zero modes with area-law entanglement scaling, and as such, break the strong thermalization hypothesis. Finally, we find zero-modes in constrained models and propose setup for observing their experimental signatures. 
\end{abstract}
\maketitle

\emph{Introduction.---}Eigenstate thermalization hypothesis (ETH)~\cite{DeutschETH,SrednickiETH} provides a specific mechanism for thermalization in isolated quantum many-body systems.  ETH suggests that the eigenstates of the Hamiltonian at a given energy density are indistinguishable by local measurements and resemble random vectors. A particular consequence of ETH is that highly excited states of quantum system feature strong entanglement. Numerical studies demonstrated that ETH can describe the vast majority of quantum systems~\cite{d2016quantum}. At the same time, possible mechanisms leading to violations of ETH are a subject of active research. Typically, ETH can be avoided due to the emergence of additional conserved quantities that may originate from special properties of Hamiltonian in integrable models~\cite{sutherland2004beautiful} or from the presence of strong disorder in the many-body localized phase~\cite{Basko06,Huse-rev,RevModPhys.91.021001}. 

While in integrable and localized systems all eigenstates disobey ETH, recently the focus shifted to systems with weak ergodicity breaking, which have a small number of weakly entangled eigenstates coexisting with the bulk of ``thermal'' eigenstates that obey ETH. These weakly entangled and thus non-thermal eigenstates were dubbed quantum many-body scars (QMBS) and were reported in a number of different models~\cite{PhysRevB.98.235155,PhysRevLett.123.147201,PhysRevResearch.1.033144,PhysRevLett.124.180604,PhysRevB.101.174308,PhysRevX.10.021051,PhysRevB.101.220305,PhysRevResearch.2.033284,PhysRevLett.119.030601}, see also Ref.~\cite{serbyn2020quantum} for a recent review. Interestingly, a large fraction of scarred systems features an exponentially large in system size nullspace which is protected by the symmetries of the model~\cite{Turner2017,PhysRevB.98.155134,PhysRevB.98.035139,PhysRevLett.123.030601}. The relevance of such nullspaces to the weak ergodicity breaking was suggested by Ref.~\cite{PhysRevLett.122.173401} which analytically constructed a particular eigenstate from the nullspace (zero mode) of so-called PXP model~\cite{PhysRevB.69.075106,PhysRevB.101.165139,PhysRevB.98.155134} as a matrix product state~(MPS). Similar zero modes were also discovered in two-dimensional models~\cite{PhysRevB.101.220304,banerjee2020quantum} and models with larger blockades~\cite{PhysRevB.103.104302}, while Ref.~\cite{PhysRevB.102.085120} proposed a systematic way of constructing parent Hamiltonian for MPS zero modes. The MPS form of these zero modes implies an area-law scaling of entanglement. Thus, such zero modes can be regarded as QMBS and provide an example of weak ergodicity breakdown. Moreover, in some cases they could be utilized as a ``vacuum'' for the construction of other QMBS states outside of the nullspace~\cite{PhysRevLett.122.173401,Ia19}. However, despite weakly entangled zero modes were established for certain models and in many-body localized systems~\cite{PhysRevResearch.2.023118}, the general conditions for their existence remain unclear.

In this work we explore the structure of the exponentially degenerate nullspaces in a large class of spin chains. We analytically construct a MPS zero mode for a broad class of two-local Hamiltonians with symmetry-protected nullspace.  For more generic Hamiltonians, we use a numerical algorithm to construct a basis in the nullspace that is ordered according to entanglement entropy~\cite{reuvers2018algorithm}. We define a notion of a least-entangled zero mode, that is shown to obey area-law entanglement scaling in a family of generic Hamiltonians, even though the majority of states in this basis features volume-law entanglement scaling, signaling thermalization~\cite{PhysRevB.98.035139}. Thus, we conjecture that all local Hamiltonians with an exponentially degenerate nullspace feature a zero mode with area-law scaling of entanglement entropy, establishing a generic route to QMBS and weak ergodicity breaking. Finally, we find the MPS zero modes in a kinetically constrained Hamiltonian and propose an experimental scheme to observe their effects in Rydberg atom arrays~\cite{bluvstein2020controlling}.

\begin{figure*}[t]
    \centering
    \includegraphics[width=0.99\linewidth]{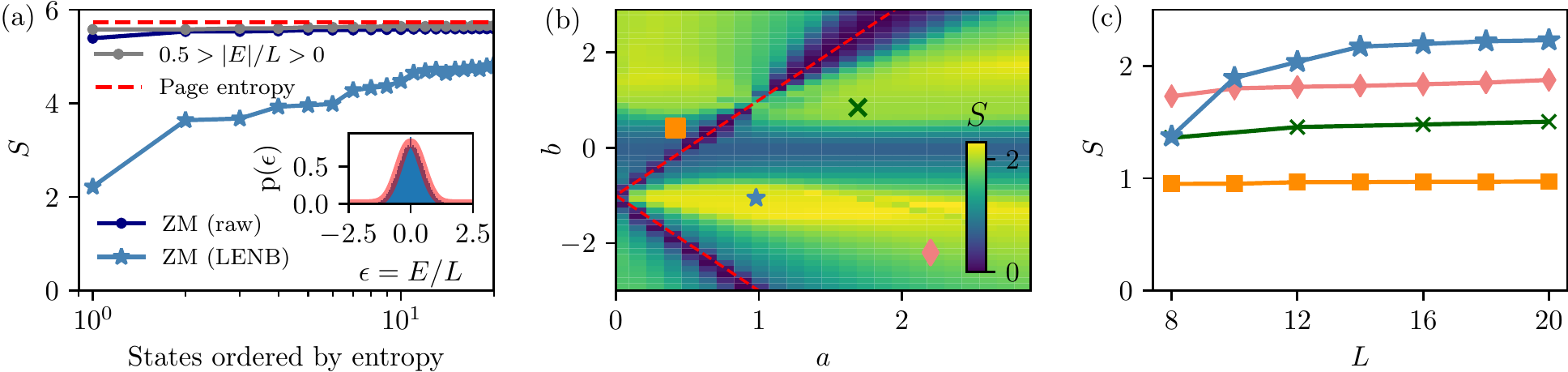}
    \caption{(a) Nullspace basis from exact diagonalization (dark blue) has uniformly high entropy similar to the entropy of the eigenstates with energy near zero (green) and close to the Page value (dashed line)~\cite{PhysRevLett.71.1291}, $S_{P} = L/2\ln2- 1/2$ . The LENB (light blue) reveals low-entangled zero modes. Inset: Density of states. Data is for $ZXZ$ model with $L = 18$ and the parameters denoted by a star in (b). (b) Entanglement of $\ket{LE_1}$ changes smoothly in the broad range of parameters of $ZXZ$ model with $L=18$. The dotted lines denote the regions where $\ket{LE_1}$ is a product state. (c)  Scaling of entanglement of the LEZM with system size $L$ is consistent with area-law. Data is shown for four different points in parameter space, corresponding to symbols of the same shape in (b).}%
    \label{fig:0}
\end{figure*}

\textit{Exponentially degenerate nullspace.---}A generic mechanism leading to the exponentially degenerate nullspace of local Hamiltonians is provided by the existence of spatial inversion symmetry and the symmetry of the many-body spectrum around zero energy~\cite{PhysRevB.98.155134,PhysRevB.98.035139}. For concreteness, we consider one-dimensional, inversion-symmetric, spin-$1/2$ chains with Hamiltonian
\begin{equation}\label{Eq:ZXZ}
H_{ZXZ} = \sum_{i=1}^L \big[X_{i} + a\left(Z_{i}X_{i+1}+ X_{i}Z_{i+1}\right)+ b Z_{i-1}X_{i}Z_{i+1}\big],
\end{equation}
parametrized by two constants, $a$ and $b$. Operators $X_i$,\,$Y_i$,\,$Z_i$ correspond to $\sigma^{x,y,z}_i$ Pauli matrices operating on the local Hilbert space of spin~$i$ spanned by $\uparrow,\downarrow$ states. We assume $L$ to be even and use periodic boundary conditions identifying spins $L+1$ and $1$. The Hamiltonian~(\ref{Eq:ZXZ}) anticommutes with the operator  $\Pi = \prod_{i}^{L} Z_{i}$,  $\{H_{ZXZ} , \Pi \} = 0$.  $\Pi$ ensures that for each eigenstate $\ket{E}$ at energy $E$, a partner eigenstate  $\Pi \ket{E} =\ket{-E}$ exists at an energy $-E$, resulting to a symmetric energy spectrum around zero energy. We note that this property holds for any Hamiltonian which contains terms with odd number of $X,Y$ operators and an arbitrary number of $Z$ operators. 

The existence of a degenerate nullspace in model~(\ref{Eq:ZXZ}) and its                         generalizations is guaranteed by the presence of spectral reflection and inversion symmetries. While this is basis-independent statement, it is easiest to understand in a computational basis, since product states with even (odd) number of $\downarrow$-spins correspond to eigenvalues $\Pi=1$ ($\Pi=-1$) respectively. Organizing basis elements into blocks with $\Pi=\pm1$, the relation $\{H_{ZXZ} , \Pi \} = 0$ implies the block-off-diagonal structure of the Hamiltonian in such basis. Presence of inversion symmetry leads to mismatch in number of basis states between different blocks provided they are restricted to a particular inversion sector, which results in a non-empty kernel. In particular, inversion-symmetric product states necessarily have even number of  $\downarrow$-spins, and there are $2^{L/2}$ such states. An explicit calculation~\cite{SOM} shows that both inversion-even and -odd sectors feature at least $2^{L/2-1}$ zero modes with even/odd number of $\downarrow$-spins. Summing up these contributions, we obtain a lower bound for the dimension of the nullspace,  $\mathcal{D}_{0} = \mathop{\rm dim}(\mathop{\rm ker} H_{ZXZ} )\geq 2^{L/2}$~\cite{PhysRevB.69.075106,PhysRevB.98.035139,SOM}.  

\textit{Analytic MPS zero mode.---}We find that all spin-1/2 Hamiltonians of the form $H = \sum_{i=1}^{L} h_{i,i+1}$, with $h_{i,i+1}$ being a two-site hermitian operator which is reflection symmetric and satisfies $\{\Pi, h_{i,i+1}\}=0$, have exact zero modes which can be represented by MPSs. As a particular example of such Hamiltonian we use~(\ref{Eq:ZXZ}) with $b=0$, while in~\cite{SOM} we discuss the more general case. 

We search for the zero-energy eigenstate of $H_{ZXZ}$ in the MPS form $\ket{\psi} = \sum_{\{a\},\{s\}} A_{a_{L} a_{1}}^{s_{1}} A_{ a_{1} a_{2}}^{s_{2}}\ldots A_{ a_{L-1} a_{L}}^{s_{L}}  \ket{s_{1}\ldots s_{L}}$, where the indices $a_i$ run from $1$ to bond dimension $\chi$, while $s_i=\uparrow,\downarrow$ labels the local Hilbert space. The local matrices $A^{\uparrow,\downarrow}$ have dimension $\chi \times \chi$. Due to the translational invariance of $H_{ZXZ}$, a sufficient condition for the state $\ket{\psi}$ to be a zero mode of the full Hamiltonian is the vanishing action of $h_{i,i+1}=(X_i+X_{i+1})/2+a(Z_{i}X_{i+1}+ X_{i}Z_{i+1})$ on the corresponding local tensors of the MPS. This condition can be written as $\sum_{s'_{i} s'_{i+1}a_{i+1}}h_{i,i+1}^{s_{i} s_{i+1},s'_{i} s'_{i+1}}A^{s'_{i}}_{a_{i}a_{i+1}}A^{s'_{i+1}}_{a_{i+1}a_{i+2}}=0$. To construct the solution for this equation we use the two dimensional  nullspace of the $h_{i,i+1}$ operator, spanned by a singlet $\ket{\uparrow \downarrow} - \ket{\downarrow\uparrow}$ and a state $\theta \ket{\uparrow \uparrow} + \ket{\downarrow\downarrow}$, where $\theta = (2a-1)/(2a+1)$ for  $H_{ZXZ}$ with $b=0$~(see~\cite{SOM} for more generic Hamiltonians).  The following choice of local matrices 
\begin{equation}\label{Eq:A}
 A ^\uparrow=  
 \begin{pmatrix}
0&1 \\
 \theta &  0
  \end{pmatrix},
\qquad 
  A^\downarrow = \begin{pmatrix}
 1&0\\
0 &  -1
 \end{pmatrix},
 \end{equation}
effectively combines these on-sites nullspaces allowing to satisfy the condition $h_{i,i+1}\ket{\psi}=0$ for any $i$, thus giving a MPS zero mode. The existence of a MPS zero mode for $H_{ZXZ}$ with $b=0$ and its two-spin generalizations that include $Y$ matrices opens the question regarding the fate of zero modes in more general Hamiltonians.

\textit{Least entangled nullspace basis.---}A systematic investigation of the nullspace is complicated due to its degeneracy and the absence of a natural basis. In order to overcome this limitation, we use entanglement to construct an unambiguous  least entangled nullspace basis (LENB), in which vectors are ordered according to bipartite entanglement entropy, $S = -\text{Tr}_{A}\rho_{A}\ln{\rho_{A}}$. The reduced density matrix $\rho_{A} = \text{Tr}_{B}\ket{\psi}\bra{\psi}$ is obtained by tracing the right half of the chain~$B$.  The LENB is constructed in an iterative procedure: first we calculate the least entangled zero mode~(LEZM), i.e.\ a superposition of all zero modes $\ket{LE_1} = \sum_{n=1}^{{\cal D}_0}c_{n}\ket{n}$, where $\sum_n |c_n|^2=1$, that has the least possible amount of entanglement. This is achieved by employing algorithms~\cite{Datta_2005,reuvers2018algorithm} which minimize the entanglement of a vector in a subspace, here chosen to be the nullspace. The resulting state, $\ket{LE_1}$ may be viewed as an analogue of the ground state in the nullspace. Once obtained, the state $\ket{LE_1}$ is projected out of the nullspace and the entanglement minimization algorithm is applied again to the remaining states resulting in $\ket{LE_2}$. The iteration of this process results in the LENB, $\{\ket{LE_{n}}\}$,\;$n$\,$=$\,$1,\ldots{\cal D}_0$, in which the states are ordered according to their bipartite entanglement entropy.

We numerically construct the LENB for the $H_{ZXZ}$ model~(\ref{Eq:ZXZ}) at generic values of parameters. Figure~\ref{fig:0}(a) shows that the $ZXZ$ model for $L=18$ chain in the inversion-symmetric sector of zero total momentum has a Gaussian density of states with a small peak at $E=0$ corresponding to nullspace. We explicitly check that the model is non-integrable and shows Wigner-Dyson level statistics~\cite{SOM}. The initial basis for the nullspace $\{\ket{n}\}$, obtained by an exact diagonalization algorithm, consists of states with approximately the same entropy, and almost coincides with the entropy of finite energy eigenstates, see Fig.~\ref{fig:0}(a). In contrast, the LENB construction results in a small number of weakly entangled states. In what follows we focus on the systematic analysis of LEZM and its entanglement scaling. 

\emph{LEZM phase diagram.---}Figure~\ref{fig:0}(b) shows the entanglement of the LEZM as a function of parameters of $ZXZ$ model~(\ref{Eq:ZXZ}) for $L = 18$. The parameters space features two special lines $b = -1\pm 2 a$ for which the  LEZM  is a product state, $\ket{\downarrow\downarrow\ldots}$ and $\ket{\uparrow\uparrow\ldots}$ respectively. These lines include the point $a=b= 1$ which corresponds to the kinetically constrained PXP model~\cite{Turner2017}. When $b = 0$, the $ZXZ$ model reduces to a sum of two-site operators for which we constructed an MPS zero mode with bond dimension $\chi = 2$ in Eq.~(\ref{Eq:A}), thus implying an area-law entanglement bounded as $S \leq \ln 2$. These  three lines in parameter space correspond to local minima in the entanglement of the LEZM as constructed by the numerical algorithm. The entropy changes smoothly around these minima which suggest the persistence of area-law entangled zero modes beyond the set of lines where analytical results are available.

We study the scaling of entropy with the system size for a wide range of model parameters in Fig.~\ref{fig:0}(c). For all simulated parameters the behavior of entanglement entropy is consistent with area-law scaling. In particular for parameters that are close to the special lines in the phase diagram entanglement does not change significantly with $L$. For other values of parameters, finite size effects are more pronounced, yet the finite size scaling is consistent with area-law and corrections decaying algebraically or exponentially with~$L$~\cite{SOM}. Crucially, for all parameters, the state $\ket{LE_1}$  is locally similar between different system sizes as witnessed by the fidelity between local density matrices, and it features a large entanglement gap in entanglement spectrum~\cite{SOM}. 

The existence of area-law entangled LEZM in a broad parameter regime in $ZXZ$ model raises the question if it is a simple consequence of the exponentially degenerate nullspace, or if the locality of the Hamiltonian is essential for its existence. This is addressed by comparing our local Hamiltonian to a random matrix Hamiltonian with similar symmetries. In this case, we observe that the least entangled zero mode follows a volume law scaling $S\propto L$, see~\cite{SOM}. In addition, we show that the distribution of the entanglement spectrum of $\ket{LE_1}$ for a random matrix approaches Marcenko-Pastur distribution~\cite{marvcenko1967distribution}. This result shows that for a random matrix the LENB construction does not lead to states which are qualitatively different from random vectors. The drastic  difference in the behavior between random matrices and local Hamiltonians implies that the area-law LEZM is related to the locality of the Hamiltonian.

\begin{figure}[t]
    \centering
    \includegraphics[width=\linewidth]{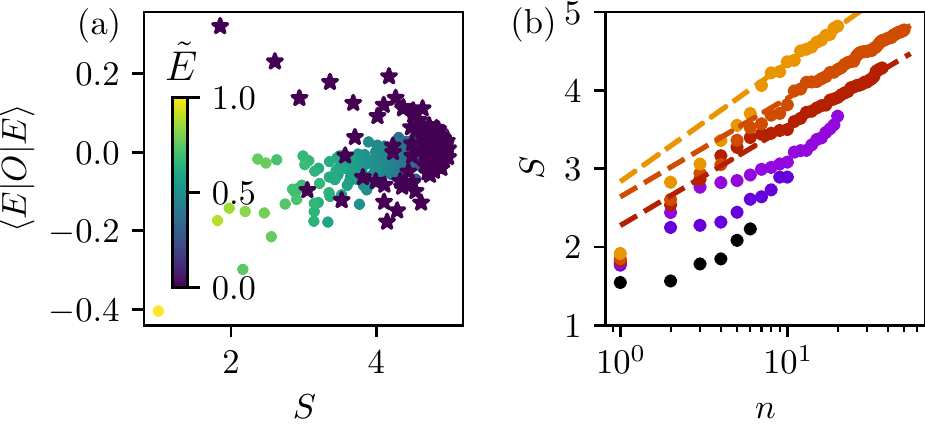}
      \caption{(a)  Diagonal matrix elements of  magnetization $O$ and eigenstates entanglement for $a= -1.3$, $b= -1.7$ and $L = 16$. The color intensity encodes the normalized energy of the eigenstates $\tilde{E} = |E/E_{0}|$, where $E_{0}$ is the ground state energy. (b)  Scaling of the entanglement entropy of the LENB with respect to the system size for the same parameters. For $L=18$ we only calculated the first 20 LEZM states.}%
    \label{fig:2}
\end{figure}

\textit{Eigenstate thermalization in LENB.---}We shift the focus from LEZM to characterizing the whole LENB from the perspective of thermalization. We compare the expectation value of the average magnetization,  $O = (1/L) \sum_{i=1}Z_{i}$  between the LENB states and non-zero energy eigenstates of $H_{ZXZ}$. ETH suggests that expectation values of the operator $O$ in eigenstates $\braket{E|O|E}$ are a smooth function of energy up to small fluctuations that are suppressed with system size~\cite{d2016quantum}. Figure~\ref{fig:2}(a) shows that as  energy $E$ approaches zero, the expectation values of observables in eigenstates concentrate around zero, while entanglement entropy is rapidly increasing. In contrast, the states from the LENB defy the expectations from ETH despite having energy $E=0$: a significant number of states from the LENB have both anomalously large expectation values of magnetization and small values of entanglement. 

To understand the difference between zero modes we explore the finite size scaling of entanglement for all LENB states, see Fig.~\ref{fig:2}(b). We observe a two component behavior, where $\ket{LE_n}$ states with small $n$ have a sub-volume law entanglement scaling, while the rest of the LENB states display  a volume law scaling. This suggests that even though the nullspace as a whole will display a thermal behaviour in agreement with previous results~\cite{PhysRevB.98.035139}, it generically hosts a small number of exceptional non-thermal states.

\textit{LEZM in $PPXPP$ model.---}We illustrate an existence of area-law zero mode in a constrained spin-1/2 model
\begin{equation}\label{Eq:PPXPP}
H_{PPXPP} =\sum^{L}_{i=1} P_{i-2}P_{i-1}X_{i}P_{i+1}P_{i+2},
\end{equation}
where  $P_{i} = (1-Z_i)/2$ is the projector to the $\downarrow$-state and we restrict to the subspace where $\uparrow$ spins are separated by at least 2 sites. This Hamiltonian corresponds to the idealized description of Rydberg atom chains with  range-2 blockade~\cite{Browaeys2020}: while in $\downarrow$ environment any given spin performs free Rabi oscillations, the presence of a nearest or next nearest neighbor $\uparrow$-spin arrests the dynamics. As the Hamiltonian has inversion symmetry and anticommutes with operator $\Pi$, it features a nullspace as described previously. The entanglement minimization in different momentum sectors reveals a number of low-entangled states in the LENB~\cite{SOM}. A particularly simple LEZM can be written analytically using two-site singlet $\ket{1}_{i} = (\ket{\uparrow \downarrow}-\ket{\downarrow\uparrow})_{i,i+1}/\sqrt{2}$ and $\ket{2}_{i} = \ket{\downarrow\downarrow}_{i,i+1}$ states stacked as $\ket{S}  = \bigotimes_{k=1}^{L/4} \left[\ket{1}_{4k} \ket{2}_{4k+2}\right]$ for $\text{mod}(L,4)=0$. The state $\ket{S}$, which was first reported in~\cite{PhysRevB.103.104302}, does not have well-defined momentum, however appropriate translations of it correspond to LEZMs in different momentum sectors~\cite{SOM}.

\begin{figure}[t]
    \centering
    \includegraphics[width=0.98\linewidth]{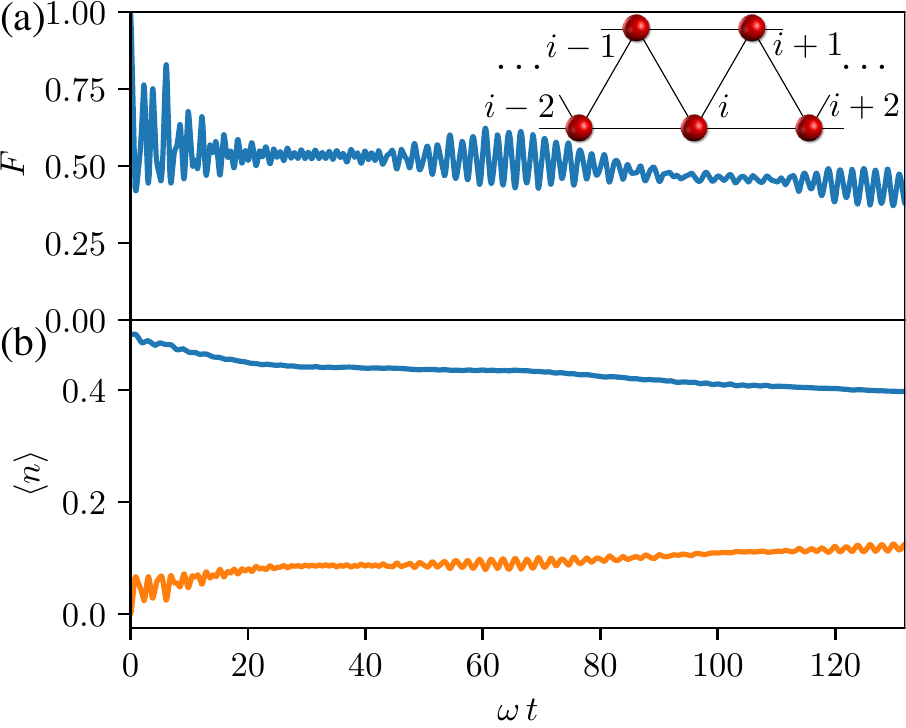}
    \caption{(a): Fidelity of the time evolved state quenching from $\ket{S}$ using the Rydberg Hamiltonian with periodic boundary conditions. (b) Dynamics of the local Rydberg densities for the two inequivalent sites of the initial state. Inset shows the ladder of Rydberg atoms and its zigzag mapping into one-dimensional chain. }%
    \label{fig:3}
\end{figure}

\textit{Experimental signatures of LEZM.---}The $PPXPP$ model can be approximately implemented by the triangular ladder of Rydberg atoms, see Fig.~\ref{fig:3} inset. Viewing the Rydberg atoms as spin-1/2 degrees of freedom, the system is governed by the Hamiltonian,
\begin{equation}\label{Eq:Ry}
H_\text{Ry} = \frac{\omega}{2}\sum^{L}_{i=1}Z_{i} + V\sum_{i\neq j} \frac{n_{i}n_{j}}{(r_{i}-r_{j})^6},
\end{equation}
where $n_i  = (1+Z_i)/2$ projects onto $\uparrow$ that corresponds to the excited state of a Rydberg atom. The long range interactions between excited Rydberg atoms decay with distance, allowing to find a nearest neighbor atom spacing $r$, such that $V_\text{nn} = V/r^6\gg \omega$. The zigzag geometry in Fig.~\ref{fig:3} inset leads to the equal strength interaction $V_\text{nn}$  between atoms $i$ and $i+1$, $i+2$. Therefore, for $V_\text{nn}\gg \omega$ the effective range-2 blockade condition emerges and the Hamiltonian~(\ref{Eq:Ry}) can be transformed into perturbed $PPXPP$ Hamiltonian using Schrieffer-Wolff transformation~\cite{SOM}. 

The perturbations to the $H_{PPXPP}$ include the spin hopping terms and also longer range interaction terms $\delta H = V_3\sum_{i=1}^L n_i n_{i+3}$. Crucially, these terms do not anticommute with $\Pi$ (since interaction terms in Eq.~(\ref{Eq:Ry}) contain only $n$, or equivalently, $Z$ operators), thus lifting degeneracy of nullspace. While the preparation of $\ket{S}$ is within the limits of current experiments in Rydberg arrays~\cite{bluvstein2020controlling}, it is an exact eigenstate only for Hamiltonian~(\ref{Eq:PPXPP}). Thus, while $\ket{S}$ remains invariant under unitary dynamics generated by $H_{PPXPP}$, we investigate its fate in time evolution under $H_\text{Ry}$ Hamiltonian.

To this end we use a Trotter-based algorithm to evolve $\ket{S}$ with $H_\text{Ry}$  for experimentally realistic value $V_\text{nn} = 2.5 \omega$, leading to a weak but yet considerable range-3 interaction terms, $V_3/ \omega=2.5/(\sqrt{3})^6\approx 0.1$ that cause splitting of the nullspace~\cite{bluvstein2020controlling}. Figure~\ref{fig:3} shows the evolution of the fidelity $ F = |\braket{S|e^{-iH_\text{Ry}t}|S)}|^2$ and local Rydberg excitation densities $\braket{n(t)}$ for a 24-atom system within the experimentally accessible timescale. We observe that although fidelity decreases from one, it remains of the order of $0.5$ even at long times. Likewise, the local densities deviate from their original values, but remain far from their equilibrium value according to the microcanonical ensemble. Such dynamics, signals that despite the presence of significant perturbations that destroy nullspace, the initialization of the system in the LEZM of idealized $PPXPP$ model results to a very slow thermalization. 

\textit{Discussion.---}We conjecture the existence of LEZM with area-law entanglement for generic local Hamiltonians with exponentially degenerate nullspace. This conjecture is supported by an analytic construction of a LEZM in the form of a MPS for a particular class of Hamiltonians, and by numerical constructions of LEZM states in the broad parameter regime. Moreover, we demonstrate the existence of LEZMs in kinetically constrained models, whose presence can be probed using Rydberg atom arrays. 

These results suggest that slightly entangled zero modes are much more common than previously thought, suggesting that the presence of a nullspace in a local Hamiltonian may be sufficient for the existence of QMBS, thus inviting the systematic studies of nullspaces. It would be interesting to understand the general conditions for the existence of a LEZM that can be represented as MPS with finite bond dimension, and extend these results to higher dimensions using projected entangled pair states representation~\cite{verstraete2008matrix}. From a numerical perspective, the existence of zero-energy eigenstates with area-law entanglement invites the development of efficient numerical algorithms based on MPS that may be able to find such states for system sizes that are beyond the reach of exact diagonalization, or even directly in thermodynamic limit. Finally, area-law entangled LEZM may be used as  ``ground states'' for creating anomalous eigenstates outside of the nullspace using local operators~\cite{PhysRevLett.122.173401,Ia19}. Understanding the conditions for a zero mode to provide a vacuum for  stable quasiparticles could result to a novel mechanism of thermalization breakdown at finite energies. 

\textit{Acknowledgments.---} We acknowledge useful discussions with V. Gritsev and A. Garkun and suggestions on implementation of $PPXPP$ model by D. Bluvstein. A.M. and M.S. were supported by European Research Council (ERC) under the European Union's Horizon 2020 research and innovation program (Grant Agreement No.~850899).

%

\onecolumngrid
\appendix
\section*{Appendix}
\section{Lower bound on number of zero modes\label{0}}
In this section we calculate the lower bound in the number of zero modes for a one-dimensional Hamiltonian with spectral reflection and spatial inversion symmetries. We focus on  even system sizes. The same calculation for odd system sizes results to a vanishing lower bound. 
 We choose $\mathcal{I}$ to be the generator of inversion symmetry and $\Pi = \prod_{i=1}^{L} \sigma^{z}_{i}$ to be the generator of spectra reflection symmetry. The following relations are true:  $[H, \mathcal{I}]=0$,  $\{H,\Pi\}=0$ and $[\Pi, \mathcal{I}] = 0$. Since the generators of the symmetries commute, we can choose the basis of the Hamiltonian to be a common eigenbasis of the two generators. Because both generators satisfy condition $\mathcal{I}^2=\Pi^2= 1$, we get 4 sectors \((N_{e +}, N_{e-}, N_{o+}, N_{o -})\) where $e/o$ denote the $\pm 1$ eigenspaces of $\mathcal{I}$ and $\pm$ denote the $\pm 1$ eigenspaces of~$\Pi$. The Hamiltonian in this basis has the form
 \begin{equation}\label{Eq:bd_Ham}
H=\begin{pmatrix}
\begin{matrix}
0 & H^{e}\\
\left(H^{e}\right)^\dag& 0
\end{matrix}
& \rvline & \bigzero \\
\hline
\bigzero & \rvline &
\begin{matrix}
0 & H^{o}\\
\left(H^{o}\right)^\dag& 0\\
\end{matrix}
\end{pmatrix}.
\end{equation}
The lower bound in the number of zero modes is given by the mismatch in the number of columns and rows of $H^{e}$ and $H^{o}$, 
\begin{equation}\label{Eq:bound}
 \mathcal{D}_{0}=\text{dim}\left(\text{ker}(H)\right)\geq |N_{e +}-N_{e -}| + |N_{o +}-N_{o -}|.
\end{equation}
To calculate this bound we evaluate the dimension of each sector. The basis states in the ($e/o$, $\pm$) sectors are written as $\ket{O_{e \pm}} = \ket{s_{\pm}} + \mathcal{I}\ket{s_{\pm}}$, $\ket{O_{o\pm}} = \ket{s_{\pm}} - \mathcal{I}\ket{s_{\pm}}$, where $\ket{s_{\pm}}$ is a computational state (and therefore an eigenstate of $\Pi$) with even/odd number of $\downarrow$-spins. To calculate the dimensions of each sector we need to take into account the set of  inversion symmetric computational states $\ket{s_{sym}}=\mathcal{I}\ket{s_{sym}}$. These states are structured as $\ket{s_{sym}} = \ket{m_{1}\ldots m_{L/2}m_{L/2}\ldots m_{1}}$ where $m_{i} = \uparrow,\downarrow$, e.g. for $4$ sites, $\ket{\uparrow \downarrow\downarrow\uparrow}$ is a symmetric state. We observe that there are $M = 2^{L/2}$ such states and they always have an even number of $\downarrow$-spins. Therefore, the dimensions of the two sectors with even number of $\downarrow$-spins are 
\begin{equation}
N_{e +}= \frac{N_{+} -  M}{2}+M, \quad N_{o +}= \frac{N_{+} -  M}{2},
\end{equation}
where $N_{+} = 2^{L-1}$ is the number of basis states with even number of $\downarrow$-spins. The two sectors with odd number of $\downarrow$-spins have equal dimensions $N_{e/o -} = \frac{N_{-}}{2}$, where $N_{-} = 2^{L-1}$  is the number of basis states with odd number of $\downarrow$-spins that coincides with $N_+$. Substituting the dimensions of all sectors to Eq.~(\ref{Eq:bound}) results to the lower bound $\mathcal{D}_{0} \geq M$. $H^{e}$ will have at least $M/2$ zero modes with an even number of $\downarrow$-spins while $H^{o}$ will have at least $M/2$ zero modes with an odd number of $\downarrow$-spins.
 

\section{Exact zero modes in two-local Hamiltonians\label{A} }
In this section we analytically calculate exact zero mode states which can be represented as a matrix product states of bond dimension $\chi = 2$ for a large class of two-local Hamiltonians including the $ZXZ$ model presented in the main text for $b = 0$. We focus on spin-1/2, translation-invariant, so-called ``two-local'' Hamiltonians $H= \sum_{i}h_{i,i+1}$ that can be written as a sum of operators acting on just two sites, for which the Hamiltonian density features spatial and spectral reflection symmetries,
\begin{equation}\label{Eq:sym}
h_{i,i+1} = h_{i+1,i},\qquad \{h_{i,i+1}, \Pi_{i}\} = 0, \qquad \Pi_{i} =Z_{i}Z_{i+1}.
\end{equation}
Due to the presence of spectral and spatial reflection symmetries the Hamiltonian density operator $h_{i,i+1}$ has two zero modes. To understand the structure of these zero modes we write the Hamiltonian density in the eigenbasis of the generator of reflections $\mathcal{I}$ and $\Pi_{i}$. Since $[\mathcal{I} , \Pi_{i}] = 0$ and  $\mathcal{I}^2= \Pi^2_{i} =\mathds{1}$, the eigenbasis is labeled by two binary quantum numbers $(i,\pi_{i})$. We use the basis consisting of triplet and singlet states with quantum numbers, $\{(1,1), \ket{\uparrow\uparrow} \}$, $\{(1,1),\ket{\downarrow\downarrow} \}$,$\{(1,-1),1/\sqrt2(\ket{\uparrow\downarrow}+\ket{\downarrow\uparrow}) \}$, $\{(-1,-1),1/\sqrt2(\ket{\uparrow\downarrow}-\ket{\downarrow\uparrow})\} $. Due to the algebraic constraints of Eq.~(\ref{Eq:sym}), the application of the Hamiltonian density to a vector with well-defined quantum numbers changes those numbers as $(i,\pi_{i})\rightarrow (i,-\pi_{i}) $. This leads to the following general structure of the Hamiltonian density in the above defined basis,
\begin{equation}
h=\left(
\begin{array}{ccc|c}
0 & 0 & c^{*}_{1} & 0 \\
0 & 0 & c^{*}_{2} & 0\\
c_{1} & c_{2} & 0 & 0\\
\hline
0 &0 &0  & 0
\end{array}
\right), \quad\text{where} \quad c_{1},c_{2} \in \mathds{C}.
\end{equation}
Solving for two  (unormalized) zero modes in different $\mathcal{I}$ sectors we get $\ket{1} = -\frac{c_{2}}{c_{1}} \ket{\uparrow\uparrow}+\ket{\downarrow\downarrow}$ and $\ket{2}=\ket{\uparrow\downarrow}-\ket{\downarrow\uparrow}$.  

To calculate the zero mode of the full Hamiltonian we propose a translation invariant matrix product state (MPS) ansatz, 
\begin{equation}\label{Eq:ham_den}
\ket{\psi} = \sum_{\{s\}, \{a\}} A^{s_{1}}_{a_{1}a_{2}}A^{s_{2}}_{a_{2}a_{3}}\ldots A^{s_{L-1}}_{a_{L-1}a_{L}} A^{s_{L}}_{a_{L}a_{1}}\ket{s_{1}\ldots s_{L}},
\end{equation}
where the local tensors $A$ have two virtual indices $a \in \{1, \chi\}$ and a physical index $s \in \{1, d\}$ with $d =2$ being  the local Hilbert space dimension. For simplicity we chose periodic boundaries  but the results also hold for open boundary conditions.  As it is mentioned in the main text, a sufficient condition for a state to be a zero mode of $H$ is that the matrix $hAA$ obtained from the action of local Hamiltonian on two sites in the MPS vanishes, 
\begin{equation}\label{Eq:con}
(hAA)^{s_{i} s_{i+1}}_{a_{i}a_{i+2}}=\sum_{s'_{i} s'_{i+1}a_{i+1}}h_{i,i+1}^{s_{i} s_{i+1},s'_{i} s'_{i+1}}A^{s'_{i}}_{a_{i}a_{i+1}}A^{s'_{i+1}}_{a_{i+1}a_{i+2}}=0.
\end{equation}
Indeed the above condition corresponds to  $h_{i,i+1}\ket{\psi} = 0$ and thus, leads to the whole MPS state being zero mode,  $H\ket{\psi} = 0$ due to translation invariance. This implies that the tensor $hAA^{(s_{i}s_{i+1})}_{a_{i}a_{i+2}}$, where the physical indices are vectorized vanishes if its matrix elements $AA_{a ,b}$ are superpositions of the zero modes of the Hamiltonian density $(\ket{1},\ket{2})$ or zero. The existence of a non-trivial solution depends on the structure of the zero mode subspace of the Hamiltonian density. For the Hamiltonian density of Eq.~(\ref{Eq:ham_den}), a solution has the form,
\begin{equation}\label{Eq:Asupp}
 A =   \begin{pmatrix}
\ket{\downarrow}&\ket{\uparrow} \\
 -\frac{c_{1}}{c_{2}}\ket{\uparrow} &  -\ket{\downarrow}
  \end{pmatrix}
\longrightarrow AA = \begin{pmatrix} -\frac{c_{1}}{c_{2}}\ket{\uparrow\uparrow}+\ket{\downarrow\downarrow} & \ket{\downarrow\uparrow} -\ket{\uparrow\downarrow}\\ -\frac{c_{1}}{c_{2}}\left(\ket{\uparrow\downarrow}-\ket{\downarrow\uparrow} \right)&  -\frac{c_{1}}{c_{2}}\ket{\uparrow\uparrow}+\ket{\downarrow\downarrow} 
\end{pmatrix}.
 \end{equation}
This means that $\ket{\psi}$  is an area-law entangled zero mode ($S \leq \ln\chi = \ln 2$). We leave the detailed analysis of this class of MPS, generalizations to higher spins,  extensions to different Hamiltonian density nullspaces and more general constraints  to future work.

\section{Entanglement minimization algorithm\label{B}}
In this section we present the numerical algorithm used to generate the least entangled nullspace basis (LENB). The algorithm is an implementation of the entanglement minimization scheme developed in~\cite{reuvers2018algorithm}, which is based on the minimization of  Renyi entropy,
\begin{equation}
S_{\alpha} = \frac{1}{1-\alpha}\ln \text{Tr}(\rho_{A}^{\alpha}),
\end{equation}
for $\alpha >1$ in a particular subspace which for our purpose is the nullspace $H_{0}$ of the Hamiltonian.  $\rho_{A} = \text{Tr}_{B} \ket{\psi}\bra{\psi}$ is the reduced density matrix of the bipartition $A = \{1,\ldots,L/2\}$, $B = \{L/2+1,\ldots, L\}$, where $L$ is the length of the spin chain. The eigenvalues $\{ \lambda_{i}^2 \}$ of the reduced density matrix  $\rho_{A}$ can be calculated by singular value decomposing the state,
\begin{equation}\label{Eq:SVD}
\text{SVD}(M) = U S V^{\dag},
\end{equation}
 using the matrix unfolding $\ket{\psi} = \sum_{I , J } M_{I J}\ket{I} \otimes \ket{J}$, where the rows/columns of the square matrix $M$ denote different basis states of the $A$/$B$ subsystems. The matrix $S = \text{diag}(\lambda_{1},\ldots,\lambda_{\text{dim}(M)})$ contains the (singular values) square roots of the eigenvalues of $\rho_{A}$ in decreasing order.

\begin{figure*}[t]
    \centering
    \includegraphics[width=1.0\linewidth]{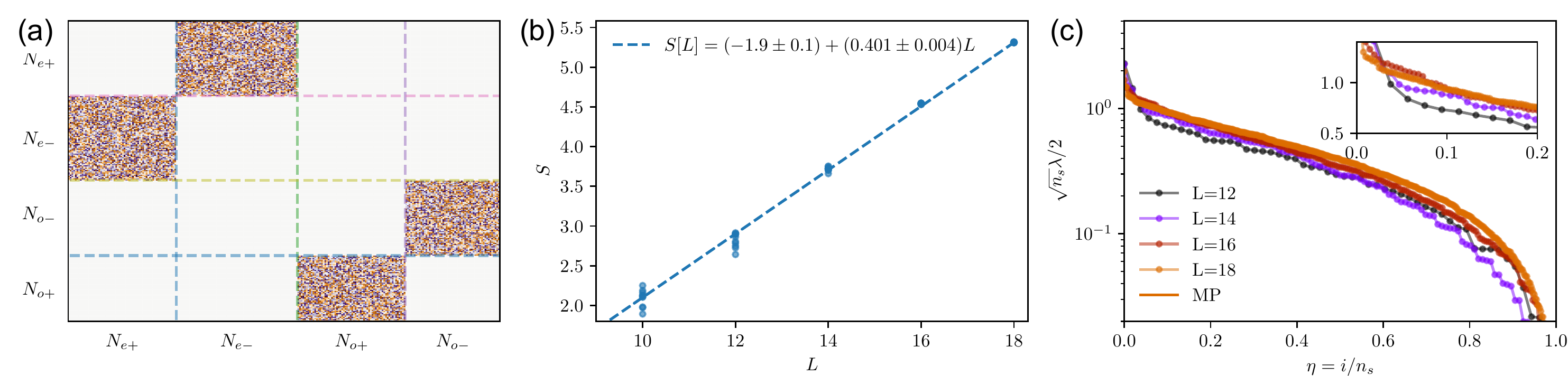}
    \caption{(a) Structure of the real valued random matrix Hamiltonian $M$. The subscripts $e/o$ denote the even/odd reflection sectors and the subscripts $\pm$ denote the eigenbase of $\Pi = \prod_{i}Z_{i}$ labeled by eigenvalue $\pm 1 $. The matrix is block diagonal with respect to reflection symmetry and block off-diagonal in the basis of $\Pi$. Matrix elements are drawn from a uniform distribution $M_{ij} \in [-1,1]$. (b) Scaling of entanglement entropy with respect to the system size. Different states in the ensemble converge to the same entropy as system size increases. (c) Distribution of the entanglement spectrum of the least entangled states for a specific random matrix realization. The axes are normalized dimension of the reduced density matrix $n_{s} = 2^{L/2}$.}%
    \label{fig:S2}
\end{figure*}
 
The algorithm for the minimization of $S_{\alpha}$ is composed of three steps~\cite{reuvers2018algorithm}:
\begin{enumerate}
\item[1] Choose a state from the subspace $\ket{0} \in H_{0}$ which consists of zero modes generated by the exact diagonalization algorithm. 

\item[2] Apply the singular value decomposition to the state, Eq.~(\ref{Eq:SVD}), and replace $S \rightarrow S'$ where $S'_{n n}  = \lambda^{2\alpha-1}_{n}$.

\item[3] Project back the state to the subspace and normalize it. Repeat steps 2-3 until entanglement entropy converges to a fixed point. 

\end{enumerate}
To minimize the  entanglement entropy ($\alpha \rightarrow 1^{+}$) we start by minimizing the Renyi entropy for some $\alpha > 1$ and slowly decrease $\alpha$ when the entanglement entropy saturates to a minimum. In particular, we find that $\alpha_{n} \rightarrow 1 + (\alpha_{n-1}-1)/2$ with an initial $\alpha_{0} = 2$, where $n \in \{ 1,N\}$ is the iteration index, provides an efficient formula to vary $\alpha$. The saturation of entanglement entropy $S_{1}$ is determined from the standard deviation $\text{SD}_{N_{loc}}(S_{1})$ over the last $N_{loc} = O(10)$ iterations. An iteration dependent threshold $\epsilon_{n} = 0.1/10^n $, is used to identify whether the entropy saturated, $\text{SD}_{N_{loc}}(S_{1}) \leq \epsilon_{n}$ . The algorithm is considered to have converged when $\text{SD}_{N_{loc}}(S_{1}) \leq10^{-4}$, which typically happens after $N=O(100)$ iterations. For large Hilbert spaces $\sim 10^4$ we fine tune the parameters to achieve optimal results  and run the algorithm for many different initial states to be sure that it converges to the global minimum. We note that since we focus on the nullspace at specific momentum sectors, the entropy will be minimized automatically for all partitions which are translations of the $A$,$B$ partition. 

To calculate the LENB we add an additional step to the algorithm: Following the entropy minimization of a state, we project it out of the subspace. Running the algorithm using the new subspace will generate a least entangled state which is orthogonal to the previous one.  Repeating this process will result to an orthonormal set of least engangled zero modes, i.e. the LENB.

\section{Least entangled zero mode of a random matrix \label{C}}

To compare the least-entangled states of the nullspace of our model to a generic random model, we  construct a random matrices with the same symmetries (besides translation symmetry) and Hilbert space dimensions as the Hamiltonian, i.e.  matrices with spatial and spectral reflection symmetry, see Figure~\ref{fig:S2}(a).  In Figure~\ref{fig:S2}(b) we show the scaling of entanglement entropy as a function of the system size for the least entangled state in the nullspace of such random matrix ensemble. We observe that the entanglement entropy scales as $S \propto L$ for every state in the ensemble. In addition,  the distribution of entropies becomes sharper as the system size increases, an indication of thermalization of the least entangled zero mode independently of the random matrix parameters. 

To further examine the structure of entanglement we  study the entanglement spectrum (i.e. the eigenvalues of the reduced density matrix) of a particular LEZM in Figure~\ref{fig:S2}(c) and compare it to the Marchenko-Pastur (MP) distribution which is the distribution of the singular values of a random matrix. We observe that the entanglement spectrum flows towards a distribution which is close to MP distribution,  further confirming the generic structure of entanglement of the least entangled zero mode of a random matrix Hamiltonian.  These results imply that the presence of weakly entangled zero modes in our systems is not just na artefact of the size of the nullspace $\sim 2^{L/2}$ and that the locality of the Hamiltonian is a critical ingredient for the existence of weakly entangled zero modes.
\begin{figure}[t]
    \centering
    \includegraphics[width=1.0\linewidth]{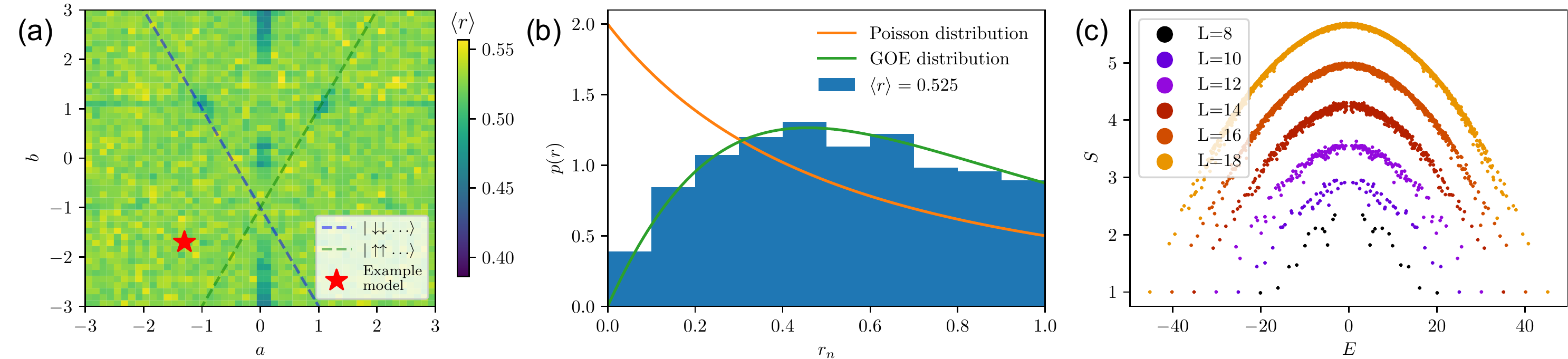}
    \caption{(a)Average level statistics $\langle r \rangle$ for the ZXZ-model for $L=18$, zero momentum and even reflection sectors agree  with the GOE prediction $\braket{r}_{GOE} \sim 0.535$. (b) Probability distribution function $P(r)$  for the parameters denoted by a red star in (a). (c) Entanglement entropies for different system sizes, for the same parameters. }%
    \label{fig:zxz}
\end{figure}

\section{$ZXZ$ Hamiltonian: Level Statistics and zero mode structure \label{D}}
In Figure~\ref{fig:zxz} we establish the chaotic nature of $H_{ZXZ} = \sum_{i=1}^L X_{i} + a\left(Z_{i}X_{i+1}+ X_{i}Z_{i+1}\right)+ b Z_{i-1}X_{i}Z_{i+1}$ studied in the main text. For this we numerically study the adjacent level statistics,
\begin{equation}
 r_n = \frac{\min{[\delta_{n+1},\delta_{n}]}}{\max{[\delta_{n+1},\delta_{n}]}}, \quad \delta_n = E_{n+1} - E_{n},
\end{equation}
which was applied by~\textcite{PhysRevB.75.155111} to quantify the breakdown of eigenstate thermalization hypothesis in strongly disordered systems and was further analyzed in~\cite{PhysRevLett.110.084101} for different statistical ensembles. For real valued chaotic Hamiltonians we expect that the level-statistics are described by those Gaussian orthogonal ensemble (GOE) while non-chaotic systems typically follow Poissonian statistics,
\begin{equation}
P_{GOE}(r) = \frac{27}{8}\frac{r +r^2}{(1+r+r^2)^{\frac{5}{2}}}, \quad \braket{r}_{GOE} \sim 0.535; \qquad P_{Poisson}(r) = \frac{2}{(1+r)^2}, \quad \braket{r}_{Poisson} \sim 0.386.
\end{equation}  
We observe that $\braket{r}\sim\braket{r}_{GOE}$  for all generic parameters $a,b$ of the $ZXZ$ Hamiltonian. We also observe that the probability distribution function is sufficiently close to the GOE prediction, showing no signs of enhancement of  $P(r)$ for small $r$ that typically stems from the absence of level repulsion and is characteristic of non-chaotic systems. In addition to the level-statistics we explore the bipartite entanglement entropy of the eigenstates of the $ZXZ$ model, Figure~\ref{fig:zxz}(c). For large enough systems, all eigenstates follow an inverse parabola which is expected from thermalizing eigenstates and a Gaussian density of states. This illustrates the uniform thermalization of the model and the absence of scarred eigenstates~\cite{Turner2017} that would be visible as ``entanglement outliers'' and are sometimes present in chaotic quantum systems.

\begin{figure}[t]
    \centering
    \includegraphics[width=0.98\linewidth]{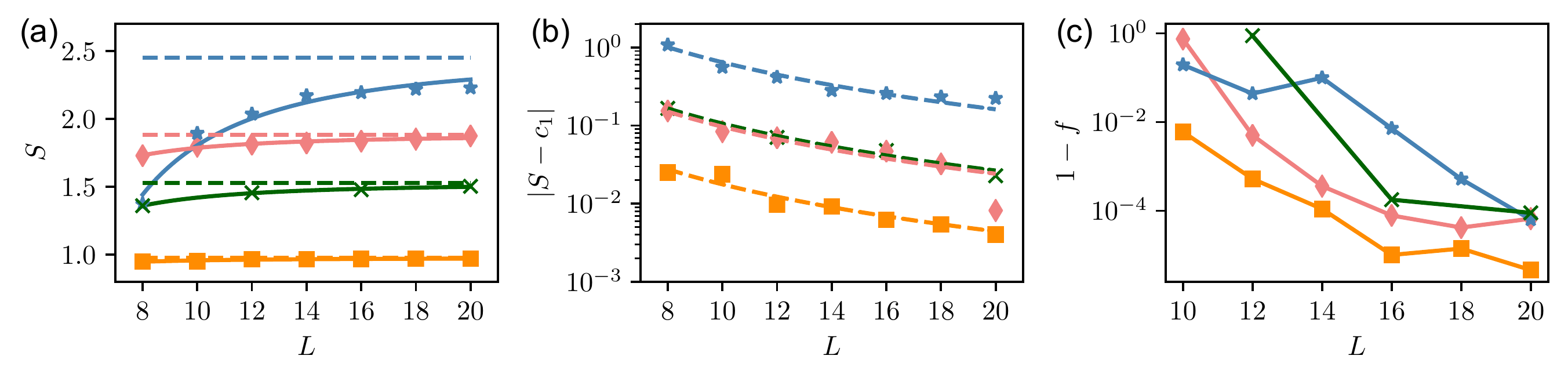}
    \caption{(a-b) Entanglement entropy fits $S = c_{1}- c_{2}/L^2$ of the LEZMs presented in the main text and behavior of $|S-c_1|$ with system size. Data is for $ZXZ$ model with parameters $(a,b)$ being equal to $(0.981,-1.0532)$ [blue stars], $(2.2,-2.2)$~[pink rhombi], $(1.692,0.84615)$~[green crosses], and $(0.4177,0.4177)$~[orange squares].
    (c) Fidelity of density matrices $f(\rho^{L-2}_{A},\rho^{L}_{A})$, for subsystems $A$ consisting of the four central sites of the chain. For the green points we show $f(\rho^{L-4}_{A},\rho^{L}_{A})$. In all cases the fidelity decreases  with the system size. }%
    \label{fig:fids}
\end{figure}

For the rest of this section we focus on the least entangled zero modes shown in Fig. 1(c) of the main text and give additional numerical evidence of the area-law scaling of the least entangled zero modes.  In Figure~\ref{fig:fids}(a) we show that the finite size corrections to the area-law are well fitted by $S(L) \sim c_{1} - c_{2}/L^{2}$.  We note however, that that exponential fit $S(L) \sim c_1+c_2 e^{-c_3 L}$ is also able to describe the saturation of entanglement. 

 In Figure~\ref{fig:fids}(c), we compare the local structure of the least entangled zero mode for different system sizes. To achieve this we construct the reduced density matrices $\rho^{L}_{A}$ of the four central sites $A = \{L/2-1,L/2,L/2+1 ,L/2+2\}$,  for spin chains of different lengths $L$. We calculate the fidelity of the density matrices for adjacent system sizes $f\left(\rho^{L-2}_{A},\rho^{L}_{A}\right)$, where $f(\sigma,\rho) = \left(\tr[\sqrt{\sqrt{\rho}\sigma \sqrt{\rho}}]\right)^2$. We observe that the fidelity approaches one as a system size increases, which implies that the algorithm converges to the same local state for all system sizes. 

\begin{figure}[t]
    \centering
    \includegraphics[width=0.98\linewidth]{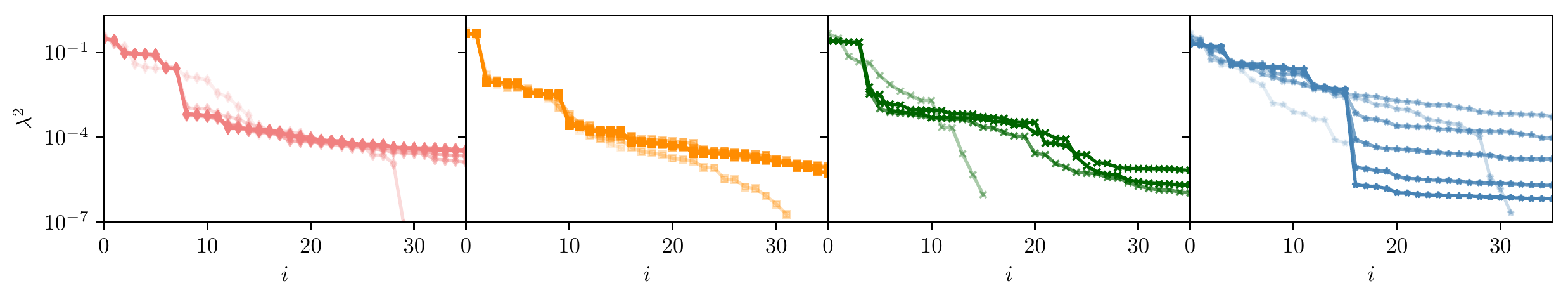}
    \caption{Largest entanglement spectra of the least entangled zero mode states shown in Fig. 1(c) of the main text. The color gradient denotes the size of the system with  $L = 8$ being the most transparent and $L = 20$ being the least transparent curves. The curves tend to collapse for large system sizes, indicating that the entanglement is generated only in the boundary between the subsystems. Moreover a jump that can be identified as ``entanglement gap'' separating large and small singular values is clearly visible in all plots.}%
    \label{fig:enspec}
\end{figure}

To further examine the fixed points of these states, we shift our comparison from local subsystems to global subsystems. In Figure~\ref{fig:enspec} we compare the different eigenspectra of the half-system reduced density matrix $\rho_{A}$ for different system sizes. We observe that for all parameters, the largest eigenvalues tend to flow towards some fixed point as we increase the system size. This is a strong indication that entanglement is only generated at the boundaries of the subsystems (area-law). 

\begin{figure}[t]
    \centering
    \includegraphics[width=0.98\linewidth]{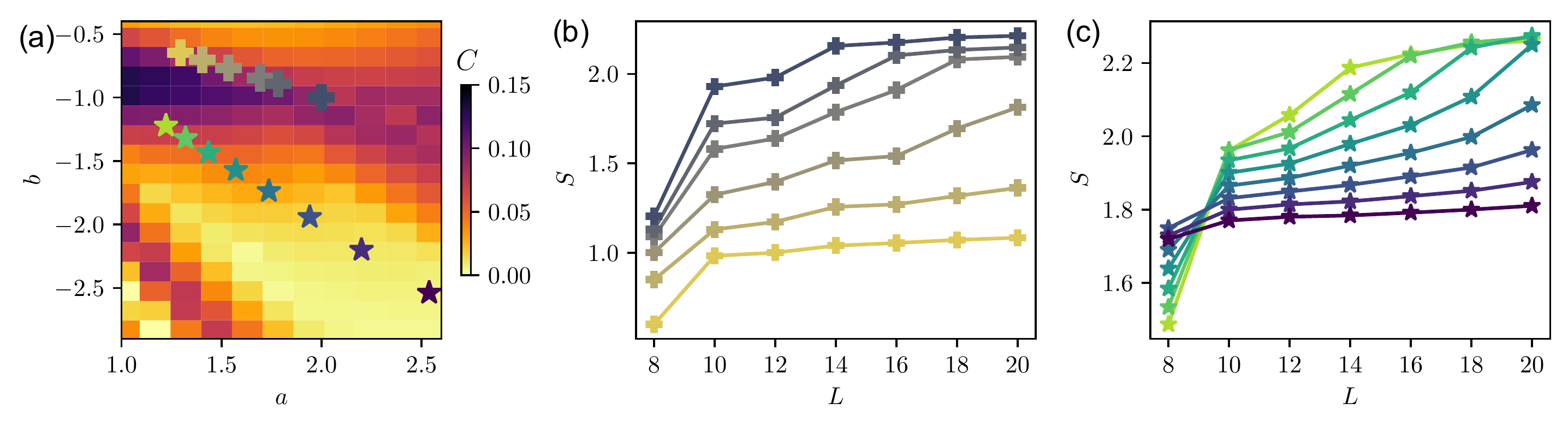}
    \caption{Entanglement scaling for different parameters. (a) The slope of entropy growth fits $ F(L)\propto C L $ for different points in the parameter space reveals fast and slow growing regimes. Crosses and star markers are points on the  $a= -2b$ and $a = -b$ lines. (b) We find that for some parameters the entropy grows, flowing towards a larger value before saturating.  In some cases the saturation may occur for system sizes beyond our numerical capabilities, thus suggesting that the growth of entropy does not rule out existence of area-law entangled zero modes. }%
    \label{fig:outliers}
\end{figure}

Finally, we have observed that for some parameter points, the entropy saturates for small system sizes and then starts increasing again, see Figure~\ref{fig:outliers}. A careful examination of the parameter regimes where this behavior is present, leads to a hypothesis that this increase of entropy is not a themodynamic feature as it leads to a second saturation point at higher entropy visible for several curves in  Fig.~\ref{fig:outliers}. 

\begin{figure}[t]
    \centering
    \includegraphics[width=0.99\linewidth]{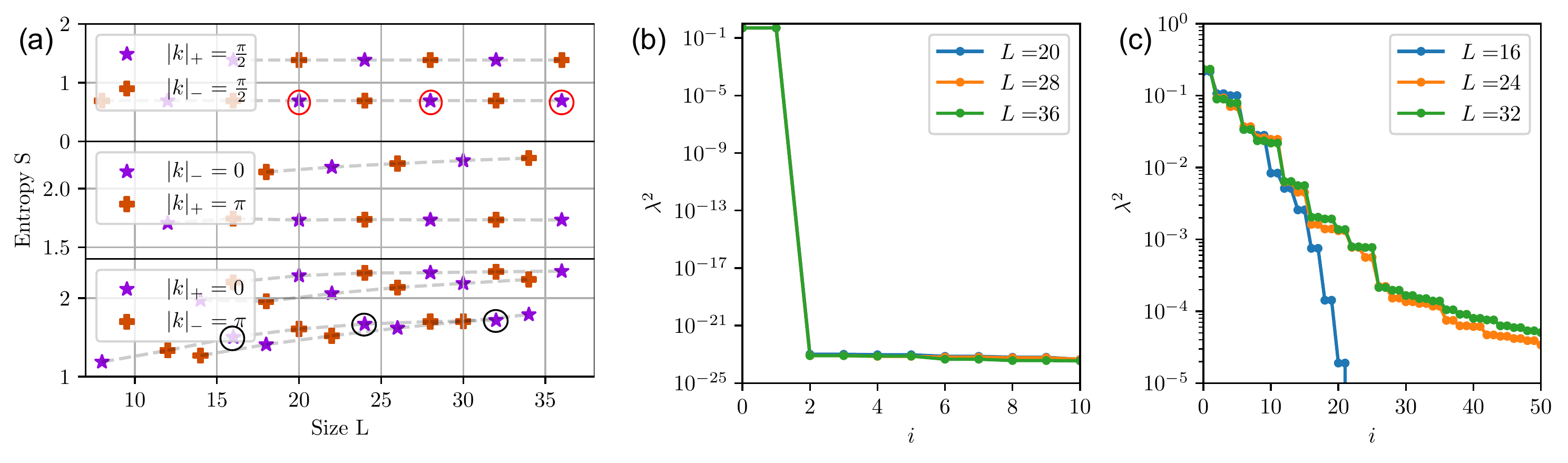}
    \caption{ Minimally entangled zero modes of the PPXPP model, Eq. (4) of the main text. (a) Entanglement of LEZMs for different momenta and reflection sectors.  Dashed lines connect points with $f(\rho^{L}_{A},\rho^{L+4}_{A})> 99 \% $ for subsystem $A$ consisting of the four central sites of the chain. (b) The entaglement spectrum of the red circled states. The algorithm accurately converges to the state defined in Eq.~(\ref{Eq:ZM}) where only two eigenvalues are finite, $\lambda^2_{1} = \lambda^2_{2} =0.5$. (c) The entanglement spectrum of the black circled states for which the density matrix is full rank converges to a fixed point distribution for large system sizes. 
    }%
    \label{fig:0modes}
\end{figure}
\section{Rydberg Hamiltonian and PPXPP model\label{E}}
In this section we derive the formal relationship between the Rydberg Hamiltonian and the PPXPP model defined in Eqs.(4)-(5) of the main text. The Rydberg blockade mechanism arises in the limit of strong nearest-neighbor interactions,$V \gg \omega$, such that the many-body Hilbert space is split into disconnected sectors distinguished by the total number of nearest-neighbor excitations. We employ Schrieffer-Wolff (SW) perturbation theory to address the connection between the Rydberg and kinetically constrained Hamiltonians ($H_{R},H_{C}$) given by Eq. (5) and Eq. (4) of the main text. The SW expansion of order $l$ corresponds to approximations of $O(\omega^{l}/V^{l-1})$.  We perform  the leading order expansion  ($l = 1$) which implies an approximation $H^{1} = H_{R}  +  O(\omega^2/V)$.

To build the expansion we split the Rydberg Hamiltonian as,
\begin{equation}
H_{R} = H^{0} + \mathcal{V}, \quad H^{0} = \frac{V}{2}\sum_{|i-j|\leq 2}n_{i}n_{j}, \quad \mathcal{V}=  \frac{\omega}{2}\sum_{i}\sigma^{x}_{i}+\frac{V}{2}\sum_{|i-j|>2} \frac{n_{i}n_{j}}{(r_{i}-r_{j})^6},
\end{equation}
where $H^{0}$ is the unperturbed Hamiltonian, $\mathcal{V}$ is the perturbation and  $(i-2,i-1,i+1,i+2)$  are the four nearest neighbors of site $i$ in the zig-zag lattice shown in Figure 3 of the main text. To derive the expansion, the perturbation is split using the generalized ladder operators, $\mathcal{V} = \sum^{M}_{m=-M}  T_{m}$, where $[H^{0} ,T_{m}] = m V T_{m}$.  The ladder operator $T_{m}$ contains all the terms of the perturbation that generate energy $m V$ when applied to an eigenstate of the unperturbed Hamiltonian. The maximum amount of unperturbed energy corresponds to $M = 4$, i.e. an atom getting excited when all four nearest neighboring atoms are already excited. The generalized ladder operators are,
\begin{equation}
T_{0} =  \frac{\omega}{2}H_{PPXPP} +\frac{V}{2}\sum_{|i-j|>2}\frac{n_{i}n_{j}}{(r_{i}-r_{j})^6}, \quad T_{m} = \frac{\omega}{2}\sum_{i}\mathcal{P}_{i}^{m}\sigma^{+}_{i} \quad \text{for}\quad m=1,\ldots,M\quad \text{with} \qquad T_{-m}= T^\dag_{m},
\end{equation}
 where the  calligraphic operators $\mathcal{P}^{m}_{i}$ are defined as projectors onto the subspace where $m$ nearest neighbors of site $i$ are simultaneously excited.  The first order Hamiltonian is obtained by rotating the Rydberg Hamiltonian using  $\mathcal{U}_{1} = \text{exp}(-\sum_{m\neq 0}\frac{T_{m}}{m V})$, 
\begin{equation}
H^{1}=\mathcal{U}^{\dag}_{1}H_{R}\mathcal{U}_{1} = H^{0} +  T_{0} +O(\omega^2/V) = H^{0} + \frac{\omega}{2}H_{PPXPP} +\frac{V}{2}\sum_{|i-j|>2} \frac{n_{i}n_{j}}{(r_{i}-r_{j})^6} +O(\omega^2/V),
\end{equation}
such that  all off-diagonal, in the unperturbed eigenbasis, elements are eliminated up to $O(\omega^2/V)$ and are ignored.  $H^{0}$ provides a constant energy shift which is proportional to the total number of adjacent Rydberg atoms being excited. In our case this is zero as we are in the subspace with no adjacent excitations. Both longer range interactions and higher order corrections break the spectral reflection symmetry and as such, the interactions cannot be too strong or too weak.

\section{LEZMs of the PPXPP model\label{F}}
Next, we focus on the LENB of the PPXPP model, $H_{PPXPP} = \sum_{i}P_{i-2}P_{i-1}X_{i}P_{i+1}P_{i+2}$. To get the LENB we fix the absolute value of momentum and spatial reflection sectors $|k|_{\pm}$ and work in the subspace with no adjacent atoms exited. We find that the LEZMs of this model also follow an area-law entanglement scaling, see Figure~\ref{fig:0modes}(a).

 In contrast to the ZXZ model, $H_{PPXPP}$  features zero modes which can be analytically calculated, and for which the half system reduced density matrix is not full rank, see Fig.~\ref{fig:0modes}(b).  These zero modes are based on the non-translation invariant exact zero mode  $\ket{S}  = \bigotimes_{k=1}^{L/4} (\ket{1}_{4k} \ket{2}_{4k+2})$, where $\ket{1}_{i} = (\ket{\uparrow \downarrow}-\ket{\downarrow\uparrow})_{i,i+1}/\sqrt{2}$ and $\ket{2}_{i} = \ket{\downarrow\downarrow}_{i,i+1}$. To illustrate the connection between the exact zero mode and the LEZMs we focus on the LEZM of $|\pi/2|_{+}$ sector as indicated by the red circles in Figure~\ref{fig:0modes}.a. This zero mode is a numerical approximation to the analytic  zero mode,
\begin{equation}\label{Eq:ZM}
\ket{S'}  = \frac{1}{\sqrt{2}}\left(\bigotimes_{k=1}^{L/4} (\ket{1}_{4k} \ket{2}_{4k+2})-\bigotimes_{k=1}^{L/4} (\ket{2}_{4k} \ket{1}_{4k+2})\right),
\end{equation}
which is a zero mode of the $k=|\pi/2|_{-}$ sector if $L = 8 n$ and $k=|\pi/2|_{+}$ sector if  $L =  8 n -4$, where $n$ is a positive integer. For this state, the reduced density matrix for the half-system bipartition has two finite eigenvalues $\lambda^2_{1} = \lambda^2_{2} =0.5$. In Figure~\ref{fig:0modes}(b) we show that the numerical algorithm converges with very high accuracy to the analytic zero mode $\ket{S'}$ which happens to be the least entangled state in that subspace. This result suggests that the entanglement minimization algorithm can also be used to find analytical zero modes as long as they are not highly entangled. In addition, the precise convergence indicates that the entanglement minimization algorithm  converges to a global minimum for all available system sizes.

 Zero modes with full rank reduced density matrices are also present in the system, see for example the black circles in Figure~\ref{fig:0modes}(b). Similarly to the ZXZ Hamiltonian we find that these zero-modes are area-law entangled and their entanglement spectrum flows towards some fixed point distribution, Figure~\ref{fig:0modes}(c).

\end{document}